\documentclass[pra,twocolumn,superscriptaddress,showpacs]{revtex4-1}
\usepackage[pdftex]{graphicx}
\usepackage{amsmath,amsfonts,amsbsy,amssymb,amsthm}
\usepackage{epstopdf}

\newcommand{\kket}[1]{| #1 \rangle\!\rangle}
\newcommand{\bbra}[1]{\langle\!\langle #1 |}
\newcommand{\bbraket}[2]{\langle\!\langle #1 | #2 \rangle\!\rangle}

\newcommand{\ket}[1]{| #1 \rangle}
\newcommand{\bra}[1]{\langle #1 |}
\newcommand{\braket}[2]{\langle #1 | #2 \rangle}
\newcommand{\matele}[3]{\langle #1 | #2 | #3 \rangle}
\newcommand{\tr}{{\rm Tr}}
\newcommand{\nn}{\nonumber\\}
\newcommand{\ii}{{\rm i}}
\newcommand{\comment}[1]{}

\usepackage{color}

\begin{document}

\title{Quantum process tomography of unitary and near-unitary maps}

\author{Charles H. Baldwin}
\affiliation{Center for Quantum Information and Control, MSC07--4220, University of New Mexico, Albuquerque, New Mexico 87131-0001, USA}
\author{Amir Kalev}
\affiliation{Center for Quantum Information and Control, MSC07--4220, University of New Mexico, Albuquerque, New Mexico 87131-0001, USA}
\author{Ivan H. Deutsch}
\affiliation{Center for Quantum Information and Control, MSC07--4220, University of New Mexico, Albuquerque, New Mexico 87131-0001, USA}

\date{\today}

\begin{abstract}
We study quantum process tomography given the prior information that the map is a unitary or close to a unitary process.  We show that a unitary map on a $d$-level system is completely characterized  by a minimal set of $d^2{+}d$  elements associated with a collection of POVMs, in contrast to the $d^4{-}d^2$ elements required for a general completely positive trace-preserving map.  To achieve this lower bound,  one must probe the map with a particular set of $d$  pure states.  We further compare the performance of  different compressed sensing algorithms used to reconstruct a near-unitary process from such data.  We find that when we have accurate prior information, an appropriate compressed sensing method reduces the required data needed for high-fidelity estimation, and different estimators applied to the same data are sensitive to different types of noise.  Compressed sensing techniques can therefore be used both as indicators of error models and to validate the use of the prior assumptions.  
\end{abstract}


\maketitle

\section{Introduction}
The development of quantum information processing devices will require new diagnostic tools for efficiently characterizing errors and  verifying performance.   While standard quantum process tomography (QPT) was initially designed in order to characterize a ``black box''~\cite{ncbook}, in practice there is often substantial prior information about the intended target map that the device is designed to implement and various diagnostic experiments initially ensure that the device is performing well.  Of particular importance is the design of devices whose target is a unitary map, e.g., quantum logic gates.  Randomized benchmarking~\cite{emerson05, knill08, magesan11} is a scalable and robust technique  that has been applied in many experiments, e.g., \cite{ryan09,chow09,olmschenk10,gaebler12,smith13a,barends14} in order to estimate the fidelity between the applied map and the target unitary.  While such information is important, particularly for evaluating whether fault-tolerant error correction is possible, in practice we would like to learn more. Given a high fidelity operation,  to further improve performance it is critical to  learn about and estimate the particular errors that led to a certain average error rate. It is therefore necessary to develop, along with benchmarking techniques, efficient QPT protocols beyond the ``black box'' model.

In the standard formulation of QPT, the applied process is an unknown completely-positive, trace preserving (CPTP) map, and therefore $d^4{-}d^2$ real numbers are required to completely characterize it~\cite{ncbook}.  If through, e.g., randomized benchmarking, we have high confidence that the applied map is close to a target unitary, we have substantial prior information. In this case, one may expect a dramatic reduction in the number of parameters, hence resources, needed for reliable estimation of the full quantum process. Such reductions have been employed in the reconstruction of near-unitary process matrices describing linear optical networks~\cite{laing12,zhou14}. Our goal in this work is to develop a general efficient protocol for QPT maps that are close to a target unitary map.  Our focus is on establishing rigorous bounds on the minimal required resources and also on procedures to validate the use of prior knowledge.

Previous workers  have studied methods to diagnose  devises that are designed to implement target unitary maps. Recently, Reich {\em  et al.} \cite{reich13} showed that by choosing specially designed sets of probe states, one can efficiently estimate the fidelity between an applied quantum process and a target unitary map.  Moreover, Gutoskia {\em  et al.}  \cite{gutoskia13} showed that the measurement of $4d^2-2d-4$ observables is sufficient to discriminate one unitary map from all other unitary maps, while identifying a unitary map from the set of all possible CPTP maps requires a measurement of $5d^2-3d-5$ observables. Utilizing prior information is important not only for efficiently gathering the required information,  but also in designing the numerical estimators that reconstruct the process from the measured data. Such techniques known as ``compressed sensing'' (CS),  originally introduced in the context of classical signal recovery \cite{davenport12}, were adapted to protocols for QPT by Kosut {\em  et al.} \cite{kosut08, shabani11} and  quantum state tomography (QST) by Gross {\em  et al.} \cite{gross10,liu11,flammia12}. These procedures reliably approximate the process (or the state) in question and  provide substantial reduction in the required data for  this task.

In this work we further study the use of prior information to perform efficient QPT of maps close to unitary evolution and integrate this with known CS  protocols based on convex optimization. Our results show two important features: (1) When we have accurate prior information, one can drastically reduce the required data needed for high-fidelity estimation. (2) Different estimators applied to the same data are sensitive to different types of noise.  These estimators can, therefore, be used as indicators of error models and to validate (to some degree) the use of prior assumptions for CS process tomography. 

After a brief review in Sec.~\ref{sec:qProc} that serves to establish our notation for the various mathematical representations of quantum processes, in Sec.~\ref{sec:tom} we describe the mathematical tools that we use for QPT.  Extending the results of Reich {\em  et al.} \cite{reich13} in Sec.~\ref{sec:unitary}, we present efficient procedures for QPT of unitary maps. In particular, we show that a given unitary map is fully characterized and completely distinguished from any other unitary map by a set of  POVMs with a total minimum of $d^2+d$ elements (measurement outcomes). This contrasts with previously known results \cite{reich13,gutoskia13}. To achieve this lower bound,  one must probe the map with a particular set of $d$ pure states, and measure the evolved states with particular POVMs.  Then, in Sec.~\ref{sec:noisy}, we study how the methods for efficient characterization of perfect unitary maps, discussed in Sec.~\ref{sec:unitary}, could be utilized in physical scenarios where the applied map is close to a unitary map by formulating the reconstruction as a convex optimization problem. We examine the behaviors of different CS estimators based on correct or faulty prior information caused by noise, and use the results as a step towards validation of prior information and as a diagnosis the nature of errors.  We present our conclusions in Sec.~\ref{sec:conc}.

\section{Review of quantum processes}\label{sec:qProc}

There are different, equivalent, ways to represent a given quantum map. Among these are the well-known, Kraus representation~\cite{kraus70}, process matrix representation~\cite{ncbook}, and Choi-Jamio{\l}kowski isomorphism~\cite{choi75}. We briefly discuss the various representations in order to establish our notation. In the Kraus representation, a completely-positive (CP) map, ${\cal E}$, is written as the sum,
\begin{equation}\label{opSum}
{\cal E}[\cdot]=\sum_{k=1}^KA_k[\cdot]A_k^\dagger,
\end{equation}
where $[\cdot]$ represents the mathematical object on which the map acts, and $\{A_k\}$ are the Kraus operators. The map is trace preserving (TP) when $\sum_k A_k^\dagger A_k=1$. The Kraus representation is not unique, and the number of independent, orthogonal, Kraus operators equals the rank of the map. If the map is unitary, ${\cal E}[\cdot]=U[\cdot]U^\dagger$, we have only one term in the above sum with  $A_1=U$. 

The process matrix representation of a quantum map  can be obtained from the Kraus representation of the map by writing the Kraus operators in a basis for complex matrices. Let $\Upsilon_n$, $n=1,\ldots,d^2$ be an orthonornomal basis for $d{\times}d$ complex matrices. Then by decompositing the Kraus operators as,
\begin{equation}\label{Ak-En}
A_k=\sum_{n=1}^{d^2} a_{nk}\Upsilon_n.
\end{equation}
we obtain the process matrix representation of the map
\begin{equation}\label{procMat}
{\cal E}[\cdot]=\sum_{n,m=1}^{d^2}\chi_{nm}\Upsilon_m[\cdot]\Upsilon_n^\dagger.
\end{equation}
The $d^2{\times}d^2$ matrix $\chi$, whose elements in the $\Upsilon_n$ basis are $\chi_{nm}$, is a the process matrix representation of the map. The CP constraint implies that the process matrix is a positive-semidefinite Hermitian matrix, $\chi\geq0$,  $\chi=\chi^\dagger$.  The process matrix of a TP map satisfies $\sum_{n,m} \chi_{nm} \Upsilon_m^{\dagger} \Upsilon_n = 1$. We say a map is pure if the process matrix is rank 1. A CP pure map is a unitary map if and only if it is a TP map.

The space of $d{\times}d$ complex matrices is a complex vector space of dimension $d^2$, with inner product defined by
\begin{equation}\label{innerP}
\bbraket{M_1}{M_2}=\tr(M_1^\dagger M_2)
\end{equation}
(where we use a `double' bra-ket notation to indicate vectors in the corresponding $d^2$-dimensional vector space).  The process matrix $\chi$ can then be seen as an operator acting on a $d^2$-dimensional complex vector space,
\begin{equation}\label{chiOp}
\chi=\sum_{n,m=1}^{d^2}\chi_{nm} \kket{\Upsilon_m}\bbra{\Upsilon_n},
\end{equation}
where $\bbraket{\Upsilon_m}{\Upsilon_n}=\delta_{n,m}$ is an orthonormal basis.  In diagonal form,
\begin{equation}\label{chiOpDiag}
\chi=\sum_{n=1}^{d^2}\lambda_{n} \kket{V_n}\bbra{V_n},
\end{equation}
with eigenvalues $\lambda_n$ and eigenvectors $\kket{V_n}$. 

Finally, the Choi-Jamio{\l}kowski representation is an isomorphism between a CP map on $d$-dimensional Hilbert space,  $\mathcal{E}[\cdot]$, and a positive-semidefinite Hermitian operator acting on  $d^2$-dimensional tensor-product Hilbert space, $\chi_{\rm c}$.  According to the isomorphism,
\begin{align}\label{choi}
\chi_{\rm c}&=\sum_{n,m=1}^{d}\ket{n}\bra{m}_A\otimes{\cal E}[\ket{n}\bra{m}_B],\nn
{\cal E}[\cdot]&=\tr_{A}\bigr\{\chi_{\rm c}\bigl([\cdot]_A^\intercal\otimes1_B\bigr)\bigr\},
\end{align}
with $\tr{({\chi_{\rm c}})}=d$, and $^\intercal$ stands for `transpose'. Being an operator on a $d^2$-dimensional Hilbert space, $\chi_{\rm c}$ has a representation as $d^2{\times}d^2$ matrix -- the Choi matrix. The elements of the Choi matrix are then given by 
\begin{equation}
(\chi_{\rm c})_{mknl}=\bra{k}{\cal E}[\ket{n}\bra{m}]\ket{l},
\end{equation}
where $m,k,n,l=1,\ldots,d$. 
The Choi matrix $\chi_{\rm c}$ is an example of a process matrix, where the operator basis $\{\Upsilon_n\}$ appearing in Eq.~(\ref{procMat}) is the `standard' basis $\{\Upsilon_n=\Upsilon_{ij}=\ket{i}\bra{j}\}$ with the relabeling of $n=1,\ldots,d^2$ by the pair $ij$ with $i,j=1,\ldots,d$.

\section{Tomography of quantum processes: Theory and numerical methods}\label{sec:tom}
QPT is the procedure by which one attempts to estimate a given process based on measurements made on the output of a well-chosen set of input states.  Assuming for a moment the ideal, unphysical case that the collected data has neither statistical errors nor systematic (SPAM) errors~\cite{gambetta12}, then an informationally complete measurement record  uniquely defines the map.  Since any representation of a general map has  $d^2(d^2-1)$ independent parameters, to specify the map one can, e.g., probe it with $d^2$ ``input'' states $\rho_j^{\rm in}$ which form a (Hermitian) operator basis. An informationally complete measurement on the output state specifies $d^2-1$ independent parameters of the map, $\rho_j^{\rm out}={\cal E}[\rho_j^{\rm in}]$.  We give two examples of such sets of states, which we will use below.  The projection onto the $d^2$ kets, 
\begin{align}\label{op basis nc}
&\ket{k},\; k=0,\ldots,d-1\nn
&\frac1{\sqrt2}(\ket{k}+\ket{n}),\; k=0,\ldots,d-2,\;n=k+1,\ldots,d-1,\nn
&\frac1{\sqrt2}(\ket{k}+\ii\ket{n}),\; k=0,\ldots,d-2,\;n=k+1,\ldots,d-1,
\end{align}
form an operator basis. These states are the generic states one typically considered in QPT protocols~\cite{ncbook}. For powers of prime dimensions, the projection onto the $d^2$ kets,
\begin{align}\label{op basis mub}
&\ket{n},\; n=0,\ldots,d-1\nn
&\ket{n;b},\; b=0,\ldots,d-1,\;n=0,\ldots,d-2,
\end{align}
also form an Hermitian operator basis. These vectors (together with $\ket{d{-}1;b}$, $b=0,\ldots,d-1$) are the elements that compose $d+1$ mutually unbiased bases (MUB), with $n$ labeling the vector in the $b$th basis~\cite{wootters89}. Using either of these sets as input states and performing an informationally complete measurement on each output state, completely specifies an arbitrary quantum map. 

Consider an arbitrary measurement defined by a POVM with elements  $\{E_{l}\geq0 \}$, with $\sum_l E_{l} = 1$.  The probability of observing an outcome $E_{l}$ for state $\rho^{\rm out}_j$,  $p_{jl} = \textrm{Tr}(\rho^{\rm out}_j E_{l})$, is expressed in terms of the process matrix using Eq.~(\ref{procMat}),
\begin{eqnarray} \label{measChi}
p_{jl} &=& \textrm{Tr} \left( \sum_{n,m=1}^{d^2}\chi_{nm}\Upsilon_m \rho_j^{\rm in} \Upsilon_n^\dagger E_{l} \right) \nonumber \\
&=&   \sum_{n,m=1}^{d^2} D_{jlmn} \chi_{mn}= \textrm{Tr} \left( D^{\dagger}_{jl} \chi \right).
\end{eqnarray}
Here $D_{jlmn} = \textrm{Tr} (\rho_j^{\rm in}\, \Upsilon^{\dagger}_n E_{l} \Upsilon_m )$, a $d^2 \times d^2$ matrix.  Alternatively, using Eq.~(\ref{choi}) we can express $p_{jl}$ in terms of the Choi matrix $\chi_{\rm c}$ as
\begin{equation} \label{measChoi}
p_{jl} = \textrm{Tr}\bigl \{ \chi_{\rm c}\bigl( (\rho_j^{\rm in})^\intercal \otimes E_{l}\bigr)\bigr\}.
\end{equation}
Comparing Eq.~(\ref{measChi}) and Eq.~(\ref{measChoi}) reveals that $D^{\dagger}_{jl} = (\rho_j^{\rm in})^\intercal \otimes E_{l}$.  Hence, the matrices $D_{jl}$ allow us to relate the Choi matrix elements directly to the measurements statistics.  

In practice, each measurement is repeated a finite number of times, and the measured data is comprised of frequencies of outcomes with finite statistical noise. We denote the frequency of the outcome $E_{l}$ in measuring the state $\rho_{j}^{\rm out}$ by $f_{jl}$.  Given the measured data, in order to characterize the quantum process in question, one must employ numerical estimators.  We consider three estimation procedures: least-squares (LS), $\ell_1$-norm CS (CS$_{\ell_1}$), and Tr-norm CS (CS$_\tr$).  The optimal solution for each of these estimators can be found using convex semidefinite programs (SDPs), given convex constraints $\chi \geq 0$ (CP constraint) and $\sum_{n,m} \chi_{nm} \Upsilon_m^{\dagger} \Upsilon_n = 1$ (TP constraint). We denote the estimated matrix returned by the procedures as $\hat\chi$. To solve the SDPs we use the MATLAB package CVX~\cite{cvx}. 

The LS estimator minimizes the (square of the) $\ell_2$-norm distance between the measurement record (consisting of frequencies) and the expected measurement record, 
\begin{align}\label{eq:ls}
\min\!{.} &\; \sum_{j,l} | f_{jl} - \textrm{Tr} (D^{\dagger}_{jl} \chi) |^2\nonumber \\
\textrm{subject to:} &\; \sum_{n,m} \chi_{nm} \Upsilon_{m}^{\dagger} \Upsilon_n = 1\nonumber \\
			&\; \chi =\chi^\dag, \, \, \chi \geq 0. 
\end{align}
The estimated matrix, $\hat\chi$, is the (constrained) maximum-likelihood solution under the assumption that the data is drawn from a Gaussian distribution. The LS program makes no prior assumptions about the nature the of process matrix we are attempting to reconstruct (besides being a CPTP process).  

The next two procedures are derived from CS in classical signal processing~\cite{candes06,candes11}.  Often, when we are attempting to estimate a process matrix, we have some prior knowledge about the physics of the device that implements the map.  Incorporating this knowledge into our numerical procedure may result in efficient estimation technique.  This, however, makes the method biased towards process matrices that satisfy our assumptions. Hence, if we made a wrong assumption about the implemented process,  the procedure can report an estimate that could have poor fidelity with the applied process. In this case the estimation procedure fails.  We will return to this problem later.  For now we discuss two main CS methods for quantum process estimation.  

The original CS estimation technique \cite{davenport12} is based on the assumption that the optimization variable (in our case, the process matrix) is a sparse vector in a known representation.  When this is the case the optimization variable can be much more efficiently extracted by minimizing its $\ell_1$-norm~\cite{candes06}.  To use this method efficiently for the problem of QPT, the process matrix should thus be close to a sparse matrix in a known basis~\cite{kosut08}. In an application such as the implementation of a quantum logic gate, one is attempting to build a target unitary map $U_{\rm t}$.   We therefore expect that  if the error in implementation is small, when expressed in an orthogonal basis $\{ V_n \}$ that includes the target process as a member, $V_1 = U_{\rm t}$, the process matrix describing the applied map will be close to a sparse matrix.  This in turn implies that the CS$_{\ell_1}$ optimization algorithm can efficiently estimate the applied process. 

We thus define the CS$_{\ell_1}$ estimator as follows:
\begin{eqnarray}
\min\!{.} &\;& \lVert {\chi} \rVert_{1} \nonumber \\
\textrm{Subject to:} 
 &\;& \sum_{j,l} |f_{jl} - \textrm{Tr} ({D}^{\dagger}_{jl} {\chi} )|^2 \leq \varepsilon \nonumber \\
&\;& \sum_{n,m} {\chi}_{nm} V_{m}^{\dagger} V_n =1\nonumber \\
&\;&\chi =\chi^\dag, \, \,\chi \geq 0, 
\end{eqnarray}
where here the matrices ${D}_{jl}$ are  represented in the $\{ V_n \}$ basis, $D_{jlmn} = \textrm{Tr} (\rho_j^{\rm in} V^{\dagger}_n E_{l}V_m )$. The first constraint equation now requires that the probabilities from our optimization variable should match our measurement frequencies up to some threshold $\varepsilon$.  The threshold is chosen based on a physical model of the statistical noise sources for the measurements.  

In the numerical simulation to be studied in Sec.~\ref{sec:noisy},  assuming a target  unitary map, $U_{\rm t}$, without loss of generality, we take the basis $\{V_n\}=\{U_{\rm t}, U_{\rm t}\Upsilon_2, U_{\rm t}\Upsilon_3, \ldots, U_{\rm t}\Upsilon_n\}$, where $\{\Upsilon_n\}$ are the generalized Gell-Mann basis.  We can regard the representation of the applied process matrix in this basis as a transformation into the ``interaction picture'' with respect to the target map; any deviation of  the applied process matrix from the projection onto $\kket{U_{\rm t}}$ indicates an error. Therefore the CS$_{\ell_1}$ directly estimates the error matrix studied in detail in~\cite{korotkov13}. This feature holds also if the target map is not a unitary map. Representing the applied map in the eigenbasis of the target map results in an error matrix; the latter is approximated by the CS$_{\ell_1}$ procedure. 

The second type of CS estimation technique we study is based on the prior assumption that the process matrix is  close to a low rank matrix.  This is equivalent to the assumption that the process is close to a (possibly unknown) unitary map.  It was shown in \cite{candes11} that one can complete a low rank matrix, $M$, for which we have only partial information by minimizing its nuclear norm, $\Vert M\Vert_* = \tr\sqrt{M^\dagger M}$.  This procedure has been applied to QST, \cite{liu11,flammia12,smith13,liu12}, for states close to pure states, and thus close to low rank density matrices, $\rho$.  Because $\rho\geq0$, $\Vert\rho\Vert_*=\Vert\rho\Vert_{\tr} = \tr (\rho)$.  We thus refer to this procedure as CS$_\tr$.

For QPT we take a similar approach.  Since our optimization variable is the process matrix $\chi$, which is positive-semidefinite, as in QST, $\Vert\chi\Vert_*= \tr{(\chi)}$.  Typically, the trace of the process matrix is part of the TP constraint equations, as for example, in the CS$_{\ell_1}$ estimator.  In the current procedure, however, we must drop any equation that constrains the trace of the process matrix. To deal with this, and maintain the maximal number of constraint equations, we take an operator basis of traceless Hermitian matrices, thereby ensuring that there is only one equation relevant to the trace of the process matrix, which is dropped as a constraint. We thus define the CS$_\tr$ estimator as follows:
\begin{eqnarray}
\min\!{.} &\;& \textrm{Tr} ( {\chi} ) \nonumber \\
\textrm{Subject to:} 
 &\;& \sum_{j,l} | f_{jl} - \textrm{Tr} ({D}^{\dagger}_{jl} {\chi} ) |^2 \leq \varepsilon \nonumber \\
&\;& \sum_{n,m \neq 1} {\chi}_{nm} \Upsilon_{m}^{\dagger} \Upsilon_n =0 \nonumber \\
&\;& \chi =\chi^\dag, \, \,\chi \geq 0,
\end{eqnarray}
where now $\chi$ and ${D}_{jl}$  are represented in a basis with $\Upsilon_1 = 1$ and the elements $\Upsilon_{m\neq1}$ are orthogonal traceless Hermitian matrices.  The sum in the second constraint include all the terms except $n=m=1$.   While the CS$_\tr$ estimator is basis independent, our numerical analysis indicates that this choice of basis improves the performance.  Also, as in the CS$_{\ell_1}$ procedure, we constrain the probabilities based on the optimization variable to match our measured frequencies up to some threshold $\varepsilon$, determined by our knowledge of the statistical noise.  Finally, the estimated process matrix, $\hat\chi$, should be renormalized such that $\tr{(\hat\chi)}=d$.


\section{Tomography of a unitary process}\label{sec:unitary}
\subsection{Minimal sets of probes and measurements}
To begin, we consider the most basic problem -- QPT of a unitary map.  Here we treat the idealized limit where the measurement record has neither statistical errors nor systematic errors.  In doing so we establish the mathematical relationships that determine the minimal set of input states and POVM elements required for perfect QPT of a unitary map.  We return to the question of statistical errors in the following subsection.

In a recent work, Reich {\em  et al.} \cite{reich13} developed an algebraic framework to identify sets of input states from which one can discriminate any two unitary maps given the corresponding output states.  In particular, a set of input states $\{\rho^{\rm in}_j\}$ provides sufficient information to discriminate  any two unitary maps if and only if the identity operator is the only operator that commutes with all $\rho^{\rm in}_j$'s in this set.  We call such set of states ``unitarily informationally complete" (UIC) set.  More generally, under unitary evolution, the output states $\{\rho^{\rm out}_j =U \rho^{\rm in}_j U^\dag\}$ completely distinguishes $U$ from any other CPTP map if and only if  $\{\rho^{\rm in}_j \}$  is a UIC set.  

An example of such a set on a $d$-level system is  
\begin{equation}
\mathcal{S}=\left\{\rho_0^{\rm in}=\sum_{n=0}^{d-1}\lambda_n \ket{n}\bra{n},\,\, \rho_1^{\rm in}=\ket{+}\bra{+}\right\},
\end{equation}
where the eigenvalues of $\rho_0^{\rm in}$ are nondegenerate,  $\{\ket{n}\}$ is an orthonormal basis for the Hilbert space, and $\ket{+} = \sum_{n=0}^{d-1} \ket{n}/\sqrt{d}$. Reich {\em  et al.} \cite{reich13} considered $\mathcal{S}$ in order to set numerical bounds on the average fidelity between a specific unitary map and a random CPTP map. In fact, $\mathcal{S}$ is a UIC set with the minimal number of input states required for complete QPT of a unitary map on a $d$-dimensional Hilbert space.  

To see this, we write the unitary map as a transformation from the orthonormal basis $\{\ket{n}\}$ to its image basis $\{\ket{u_n}\}$,
\begin{equation}\label{unitary}
U=\sum_{n=0}^{d-1}\ket{u_n}\bra{n}.
\end{equation}
In its essence, the task in QPT of a unitary map is to fully characterize the basis $\{\ket{u_n}\}$, along with the relative phases of the summands $\{\ket{u_n}\bra{n}\}$. 
By probing the map with $\rho_0^{\rm in}$, we obtain the output state $\rho_0^{\rm out}=U\rho_0^{\rm in}U^\dagger=\sum_{n=0}^{d-1}\lambda_n \ket{u_n}\bra{u_n}$, which we fully characterize by means of an informationally complete POVM. Such a POVM has at least $d^2-1$ elements. We then diagonalize $\rho_0^{\rm out}$ and learn $\{\ket{u_n}\bra{u_n}\}$. Without loss of generality, we take the global phase of $\ket{u_0}$ to be zero. Next, we probe the map with $\rho_1^{\rm in}$, and fully characterize the output state  $\rho_1^{\rm out}=\sum_{n,m=0}^{d-1} \ket{u_n}\bra{u_m}/d$ with an informationally complete POVM. The $\{\ket{u_n}\}$ are calculated according to the relation, $\ket{u_n}\bra{u_n}\rho_1^{\rm out}\ket{u_0}=\ket{u_n}/d$. This procedure identifies a unique orthonormal basis $\{\ket{u_n}\}$ if and only if the map is a unitary map.

So far we have not specified the nature of the ``informationally complete measurement on the output state.'' Under a unitary evolution $\rho_1^{\rm out}$ is a pure state.  To fully characterize it, we may therefore use a measurement that is informationally complete for pure states. Flammia {\em et al.}~\cite{flammia05} showed that the $2d$ operators,
\begin{align}\label{psic mixed}
&E_0=a\ket{0}\bra{0},\nn
&E_n=b(1+\ket{0}\bra{n}+\ket{n}\bra{0}),\; \;n=1,\ldots,d-1,\nn
&\widetilde{E}_n=b[1+\ii(\ket{0}\bra{n}-\ket{n}\bra{0})],\; \;n=1,\ldots,d-1,\nn
&E_{2d}=1-\Bigl[E_0 +\sum_{n=1}^{d-1}(E_n+\widetilde{E}_n)\Bigr]
\end{align}
with $a$ and $b$ chosen such that $E_{2d}\geq0$, represent an informationally complete POVM for pure states with the minimal number of outcomes. That is, all pure states in a $d$-dimensional Hilbert space (except ones in a set of measure zero \cite{footnote1}) are completely determined by the probabilities of these POVM elements. The total minimal number of POVM elements that are needed for complete characterization of a unitary map using the above procedure is thus, $d^2-1+2d$. 

While $\mathcal{S}$ is the minimal UIC set, in practice we do not have reliable procedures to produce a desired,  reproducible, mixed state $\rho^{\rm in}_0$.  We thus turn our attention to minimal UIC sets that are composed only of pure states (arbitrary pure states can be reliably produced using the tools of quantum control~\cite{smith13a}). Such UIC sets are composed of $d$ pure states that form a nonorthogonal vector basis for the $d$-dimensional Hilbert space. For example, the set
\begin{align}\label{n+}
\ket{\psi_n}&=\ket{n},\; n=0,\ldots,d-2,\nn
\ket{\psi_{d{-}1}}&=\ket{+}=\frac1{\sqrt d}\sum_{n=0}^{d-1}\ket{n},
\end{align}
 a subset of the states of Eq.~(\ref{op basis mub}), is a minimal UIC set of pure states. A similar set (with $d+1$ elements) was considered in~\cite{reich13}. Here we focus on a different set of $d$ pure states that is UIC,
\begin{align}\label{0+n}
\ket{\psi_0}&=\ket{0}\nn
\ket{\psi_n}&=\frac1{\sqrt2}(\ket{0}+\ket{n})\; n=1,\ldots,d-1.
\end{align}
This is a subset of the standard states used in QPT, Eq.~(\ref{op basis nc}). The only operator that commutes with all of the projectors $\{\ket{\psi_n}\bra{\psi_n}\}$, $n=0,\ldots,d-1$, is the identity operator.

A generic tomographic procedure for a unitary process using the set given in Eq.~(\ref{0+n}) can be described as follows.  First let the  map act on $\ket{\psi_0}$ and make an informationally complete measurement on the output state $U\ket{\psi_0}=\ket{u_0}$, from which we can obtain the state $\ket{u_0}$ (up to a global phase that we can set to zero). Next, let the unitary map act on $\ket{\psi_1}$ and perform an informationally complete measurement on the output state $U\ket{\psi_1}\bra{\psi_1}U^\dagger$. From the relation $U\ket{\psi_1}\bra{\psi_1}U^\dagger\ket{u_0}=\frac1{2}(\ket{u_0}+\ket{u_1})$ we obtain the state $\ket{u_1}$, including its phase relative to $\ket{u_0}$. We repeat this procedure for every state $\ket{\psi_n}$ with $n{=}1,{\ldots},d{-}1$, thereby obtaining all the information about the basis $\{\ket{u_n}\}$, including the relative phases in the sum of Eq.~(\ref{unitary}), and completing the tomography procedure for a unitary map.  In this protocol, because input and output states are ideally pure, we may use the POVM of Eq.~(\ref{psic mixed}), which has $2d$ measurement outcomes, as the informationally complete measurement on each output state. We thus conclude that the total number of POVM elements needed for reconstruction of a unitary map based on the protocol above is $d\times2d=2d^2$. This is the minimal number of measurement outcomes required to distinguish a unitary map from an arbitrary CPTP map, when the map in probed with  pure input states.

While $2d^2 > d^2-1+2d$, the total number of measurement outcomes required for the minimal (mixed) UIC set $\mathcal{S}$, by using a UIC set of $d$ pure states, we can fully distinguish a unitary map from the set of all unitary maps by measuring sets of POVMs that have in total only $d^2+d$ elements.  The key ingredient is to note that in the  procedure above, we have not used the orthogonality of the basis $\{\ket{u_n}\}$. By taking this into account, we can reduce the number of required  measurement outcomes on each output state. The first step is, as before, to use $\ket{\psi_0}$ as a probe state, and perform an informationally complete measurement, which has $2d$ outcomes, on the output  state,  $\ket{u_0}$. From this measurement we determine  $\ket{u_0}$ completely (up to an irrelevant global phase). Note, the probability to detect the outcome $E_0$ is 
\begin{equation}\label{c00}
p_{00}=\matele{u_0}{E_0}{u_0}=a|c_{00}|^2, 
\end{equation}
and since the amplitude $c_{00}$ can be taken to be positive without loss of generality, we deduce that $c_{00}=\sqrt{p_{00}/a}$.  Similarly, from the probabilities of the outcomes $E_n$ and $\widetilde{E}_n$ we obtain the real and imaginary part of $c_{0n}$, respectively, 
\begin{align}\label{c0n}
p_{0n}&=\matele{u_0}{E_n}{u_0}=b[1+2c_{00}{\rm Re}(c_{0n})],\nn
\tilde{p}_{0n}&=\matele{u_0}{\widetilde{E}_n}{u_0}=b[1+2c_{00}{\rm Im}(c_{0n})].
\end{align}
Thus, we obtain full information about the state $\ket{u_0}$. This procedure fails for states with $c_{00}=0$, a set of measure zero.  Next, we  probe the unitary map with $\ket{\psi_1}$ of Eq.~(\ref{0+n}), and perform an informationally complete measurement on the output state,  $\frac1{\sqrt2}(\ket{u_0}+\ket{u_1})$. However, since $\ket{u_1}$ is orthogonal to $\ket{u_0}$, it is sufficient to make a measurement that yields only the first $d-1$ probability amplitudes $c_{1n}=\braket{n}{u_1}$, $n=0,\ldots,d-2$ and then use the orthogonality condition $\braket{u_0}{u_1}=\braket{u_1}{u_0}=0$ to calculate the $d$th amplitude,   $c_{1d{-}1}$. A measurement with $2d-2$ outcomes can be, for example, the measurement of Eq.~(\ref{psic mixed}), but with  $n=0,\ldots,d-2$. Therefore, to measure the state $\ket{u_k}$, $k=0,\ldots,d-1$ we perform a measurement with $2d-2k$ outcomes, and use $2k$ orthogonality relations. 

Overall, if we have prior information that the map is unitary, this protocol shows that we can reconstruct it from  $d^2+d$ measurements outcomes. This number is, by construction, the minimum number of POVMs elements required to reconstruct a unitary map.  Though this protocol is slightly more efficient than the previous one, it assumes much more. Crucially, the former protocol, assumes {\em no prior knowledge of the map}, and can be cast as a convex optimization program to uniquely reconstruct an applied unitary map after informationally complete measurements on $d$ input pure state. This features makes it useful for laboratory implementation, when estimating an applied map from measured data. 

\subsection{Simulating tomography of a unitary processes}
To study the behavior of the reconstruction using the UIC sets of pure states discussed above, in this section we simulate QPT based on the LS estimator, Eq.~(\ref{eq:ls}). Here, and throughout, we consider a target unitary map on a qudit of dimension $d=5$, a Hilbert space dimension sufficient to test the the general performance of the estimators. We  generate a random process matrix  and let it act on a preassigned set of input states. We then use an informationally complete measurement on the output states to generate the expected probability distribution according to Eq.~(\ref{measChoi}). Finally, we simulate the measurement record by the outcome probabilities plus a Gaussian distributed random variable $W_{jl}$ with mean zero and variance one,
\begin{equation} \label{freq}
f_{jl} = p_{jl} + \sigma W_{jl}.
\end{equation}
The magnitude $\sigma$ is held constant for all $j$ and $l$.  In this way, by repeating the measurement many times, the frequencies are normal Gaussian distributed random variables with mean given by the probability $p_{jl}$ and variance describing the noise of large but finite counting statistics.

To evaluate the performance of the estimators, we calculate the (Uhlmann) fidelity between the reconstructed process matrix, $\hat{\chi}$ and the process matrix of a target unitary map, $\chi_{\rm t}=\kket{U_{\rm t}}\bbra{U_{\rm t}}$,
\begin{equation}\label{fidelity}
F(\hat{\chi}, \chi_{\rm t}) =\frac1{d^2}\left({\rm Tr}\sqrt{\sqrt{\chi_{\rm t}}\hat{\chi} \sqrt{\chi_{\rm t}}}\right)^2 =\frac1{d^2} \bbra{U_{\rm t}}\hat{\chi}\kket{U_{\rm t}},
\end{equation}
as a function of the number of input states. By this we mean that we simulate an estimator that uses all of the data derived by inputing the first $k\leq d$ states in a particular order, and   measuring  the output states with an informationally complete POVM. The input states are taken to be those of Eqs.~(\ref{op basis nc}) and (\ref{op basis mub}).  As we are considering the ideal case of a perfect unitary map, each output state is measured with the informationally complete measurement for pure states of Eq.~(\ref{psic mixed}).  

\begin{figure}[t]
\centering
\includegraphics[width=.95\linewidth]{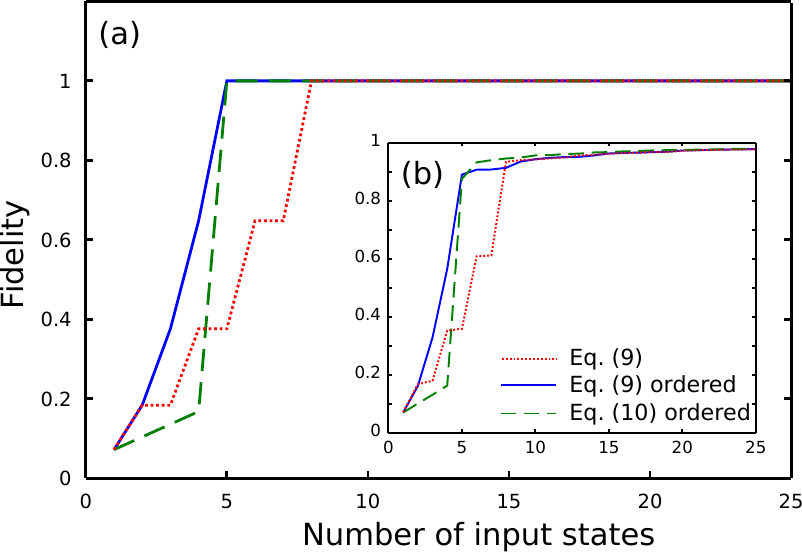}
\caption{Fidelity between  a unitary map on a $d=5$ dimensional Hilbert space and the LS estimate of the process matrix (averaged over 50 Haar-random unitary maps) as a function of the number of input states in the absence of statistical noise (a), and  in the presence of statistical noise (b).  The data represented in the dotted (red) line is obtained from an informationally complete measurement on the states produced by applying the unitary map to all $d^2$ input states in the order specified in Eq.~(\ref{op basis nc}). In the solid (blue) line the first five input states are of the ones of Eq.~(\ref{0+n}),  and the remaining states are chosen from Eq.~(\ref{op basis nc}) in an arbitrary order.  In the dashed (green) line the first five input states are of the ones of Eq.~(\ref{n+}) and the remaining state from Eq.~(\ref{op basis mub}).  Unit fidelity is obtained after $d=5$ input states in a UIC set. In the presence of statistical error, as seen in  (b), the main features of the fidelity remain qualitative the same -- by choosing the correct set of input states we obtain a reliable reconstruction after $d$ input states, up to statistical error. The statistical fluctuations in the data is modeled by Eq.~(\ref{freq}),  a normal distribution, with $\sigma=10^{-4}$.}
\label{fig:unitary_diff_sets}
\end{figure}

To begin, we treat the idealized situation where there is no noise in the data, i.e., $\sigma=0$ in Eq.~(\ref{freq}).  We clearly see in Fig.~\ref{fig:unitary_diff_sets}a that the order in which one uses the input states affects the way in which we gain information about the process. In the dotted (red) line we use the generic order of input states used for general process tomography given in Eq.~(\ref{op basis nc})  starting with $k=0$ and running over $n$. The plateaus in the plot indicate that we gain information only from particular states in that set, while others do not provide additional information. Therefore, to obtain an efficient reconstruction of a unitary map with $d=5$ pure input states, the latter must form a UIC set. Such states are used in the solid (blue) line, and the dashed (green) line in  Fig.~\ref{fig:unitary_diff_sets}. In the solid line we input the first $d=5$ input states in Eq.~(\ref{0+n}) and then the remainder of the states of Eq.~(\ref{op basis nc}).  In the dashed line, on the other hand, the first five input states are of the ones of Eq.~(\ref{n+}), followed by the remainder of the states of Eq.~(\ref{op basis mub}). Note that in our LS program, for the reconstruction of the process matrix, we did not assume any structure on the reconstructed map except that it is a CPTP map. Yet, after $d$ input states of Eq.~(\ref{0+n}) or  Eq.~(\ref{n+}), the LS estimator returns the target map. Thus, $d$ input states are sufficient to completely identify a unitary map among {\em all} CPTP maps. 

Because we assumed a (nonphysical) hypothetical situation where there is no noise in the data, the fidelity between the reconstructed map and the target map eventually reaches unity. In any physical implementation, however, even in the absence of any systematic errors, finite sampling always results in statistical noise. Qualitatively, such noise does not  affect the key features of the results above in the large sampling (Gaussian noise) limit, as seen in Fig.~\ref{fig:unitary_diff_sets}b. The effect of the statistical noise on the fidelity between the estimate and the target map is two-fold: it reduces its maximal value below one (depending on the level of noise), and after $d$ input states, the plateau in the fidelity acquires a shallow slope.   

While the addition of some statistical noise to the measurement record still allows us to obtain, more or less, all the information about the process with $d$ input states, if the map itself is far from unitary, the reconstruction can fail dramatically.  We address this behavior and methods to validate prior assumptions in the next section.

\section{Tomography of a near-unitary process}\label{sec:noisy}
While the previous section established the behavior of the LS estimator in the situation that the applied map was equal to the target unitary map, in any physical implementation this is never exactly true.  Our goal is to understand how well the idealized case considered above carries over to the realistic case under the assumption that the errors in implementation are sufficiently small. Moreover, we seek to {\em validate} the prior assumptions that are used in the protocol, and to use the estimation procedures to {\em characterize} the kinds of errors that lead to imperfections in the applied map.

Let us denote the target unitary process ${\cal E}_{\rm t}$ with corresponding process matrix $\chi_{\rm t}$.   Due to experimental imperfections, the process actually being implemented is ${\cal E}_{\rm a}$, with corresponding process matrix $\chi_{\rm a}$.  We assume good experimental control so that the implementation errors are low, hence, $\chi_{\rm a}$ is close to $\chi_{\rm t}$. Our goal is to reconstruct $\chi_{\rm a}$ as faithfully as possible in order to diagnose the errors that led to the creation of ${\cal E}_{\rm a}$. 

We consider two types of imperfections in the implementation of the map: ``coherent'' errors and ``incoherent'' errors.  A coherent error is one where the applied map is also unitary, but ``rotated'' from the target. We define incoherent errors as errors that are not coherent errors, for example, statistical mixtures of different unitary maps arising from inhomogeneous control or decoherence. 

Under the assumption that the applied process is close to a known target unitary map, we expect the resource reduction obtained in the ideal case, where there were no implementation errors, to carry over here to the noisy case.  We thus study the performance of the estimators probed with states that are efficient for reconstruction of a unitary map, e.g., the states of Eq.~(\ref{0+n}). Moreover,  using the knowledge that $\chi_{\rm a}$ is close to $\chi_{\rm t}$,  we expect that in the eigenbasis of $\chi_{\rm t}$,  $\chi_{\rm a}$ is close to a sparse matrix, and we therefore expect the CS$_{\ell_1}$ procedure to yield an estimate $\hat\chi$ with high fidelity with $\chi_{\rm a}$ with very little data.  In addition, since the CS$_\tr$ estimator performs well when estimating process matrices that are close to low rank matrices, when the target map $\chi_{\rm t}$ is a unitary map, and $\chi_{\rm a}$ is close to $\chi_{\rm t}$, we expect the CS$_\tr$ estimator should return an estimate with high fidelity to $\chi_{\rm a}$ after input of $d$ UIC states. 

In some cases, however, our assumption that  $\chi_{\rm a}$ is close to $\chi_{\rm t}$ is flawed.  Then these estimation procedures will in general return an estimate $\hat\chi$ that is very different from $\chi_{\rm a}$.  In what follows we will see that one can utilize the sensitivity of the estimators to the prior assumptions as an indicator for the possible types of errors that occurred in the implementation of the map, and to validate, to some degree, whether prior assumptions are justified.

As in Sec.~\ref{sec:unitary},  in the numerical simulations below we consider the estimation of random maps on a $d=5$ dimensional Hilbert space.  We ignore SPAM errors, assuming that the input states are perfectly prepared and that the output states are perfectly measured (up to statistical errors) with informationally complete measurement of the MUB.  The target map, ${\cal E}_{\rm t}$ is a Haar-random unitary map, and the applied map is generated by its composition with a map that describes the error channel, ${\cal E}_{\rm a} = {\cal E}_{\rm err}\circ{\cal E}_{\rm t}$. For coherent errors, we take ${\cal E}_{\rm t}[\cdot]=U_{\rm t}[\cdot]U_{\rm t}^\dagger$ and ${\cal E}_{\rm err}[\cdot]=U_{\rm err}[\cdot]U_{\rm err}^\dagger$, where $U_{\rm err}=e^{\ii\eta H}$, with $\eta\geq0$, and $H$ is a random Hermitian matrix selected by the Hilbert-Schmidt measure. For incoherent errors we consider the case where 
\begin{equation}
{\cal E}_{\rm err}[\cdot]=(1-\xi)[\cdot]+\xi\sum_{n=1}^{d^2}A_n[\cdot]A_n^\dagger ,
\end{equation}
so that the applied map is given by 
\begin{equation}\label{depCh}
{\cal E}_{\rm a}[\cdot]=(1-\xi)U_{\rm t}[\cdot]U_{\rm t}^\dagger+\xi\sum_{n=1}^{d^2}A_nU_{\rm t}[\cdot]U_{\rm t}^\dagger A_n^\dagger.
\end{equation}
The set $\{A_nU_{\rm t}\}$ are Kraus operators associated with a CP map.  Eq.~(\ref{depCh}) can be interpreted as mixture of two maps --  the target map and an error-related map, with a mixing parameter $\xi\in[0,1]$. The $\{A_n\}$'s are generated by choosing a Haar-random a unitary matrix $U$ of dimension $d^3$, and a random pure state of dimension $d^2$ from the Hilbert-Schmidt measure, $\ket{\nu}$, such that $ A_n= \bra{n}U\ket{\nu}$ where the set $\{\ket{n}\}$ is a computational basis~\cite{bruzda09}.  

\begin{figure}[t]
\centering
\includegraphics[width=1\linewidth]{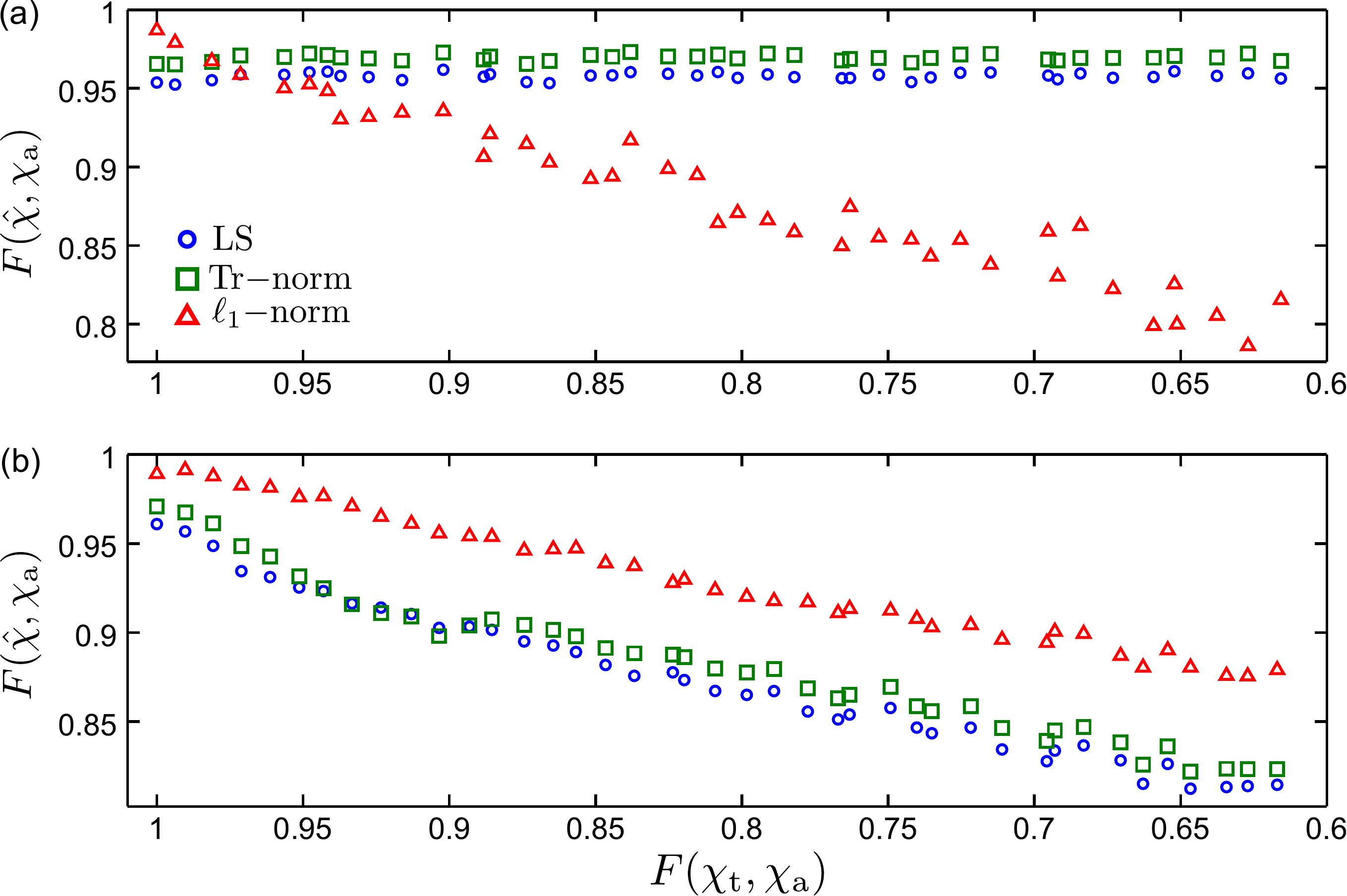} %
\caption{The fidelity between the estimate and the applied map as a function of the fidelity between the target map and the applied map for the case of coherent errors (a) and incoherent errors (b). The estimates are obtained with data from only the five UIC input states of Eq.~(\ref{0+n}). Each data point in the plots is obtained by an average over 20 random target unitary maps each with a random error map. (a) Coherent errors: The applied map is given by $U_{\rm a}=U_{\rm err}U_{\rm t}$, where $U_{\rm t}$ is a Haar-random unitary map, $U_{\rm err}=e^{\ii\eta H}$, $\eta\in[0,3]$, and $H$ is a random Hermitian matrix selected by the Hilbert-Schmidt measure, normalized to $\tr{H}=1$.  While the fidelities of the estimator based on the CS$_ {\tr}$ and the LS minimizations remain more or less constant, the fidelity of the estimator based on the CS$_ {\ell_1}$ minimization decreases as the fidelity between the applied and the target maps decreases. This is an indicator of the sensitivity of CS$_ {\ell_1}$ to coherent errors. (b) Incoherent errors: The applied map is given by Eq.~(\ref{depCh}). The numerical simulation was done by choosing at random values of $\xi$ from a uniform distribution on [0,0.6]. For each value we generated Haar-random unitary target maps $U_{\rm t}$, and randomly selected 25 Kraus operators from the Hilbert-Schmidt measure as prescribed in Ref.~\cite{bruzda09}. The estimate based on the CS$_ {\ell_1}$ minimization performs better  than the   estimate based on the CS$_ {\tr}$ and the LS minimization procedure, and thus, the  CS$_ {\ell_1}$ estimator is more robust to incoherent errors of the form of Eq.~(\ref{depCh}) than either the CS$_ {\tr}$  or the LS estimator.} 
 \label{fig:fidFid}
\end{figure}

We first test the sensitivity of the CS$_{\ell_1}$,  CS$_\tr$, and LS estimators to the error type and magnitude. In Fig.~\ref{fig:fidFid} we plot the fidelity between the applied matrix, $\chi_{\rm a}$, and the reconstructed matrices, $\hat\chi$, determined by  each of the three estimators, as a function of the fidelity between the applied process $\chi_{\rm a}$ and the target $\chi_{\rm t}$.   The latter fidelity, $F(\chi_{\rm t},\chi_{\rm a})$,  is a measure of the magnitude of the error in the applied process.  Each data point is obtained by an average of 20 random error processes (Fig.~\ref{fig:fidFid}a for coherent errors and Fig.~\ref{fig:fidFid}b for incoherent errors) based on informationally complete measurements from $d=5$ UIC input states.  As expected, all of the estimators return reconstructions that have high fidelity with the applied map when the applied map is close to the target unitary map ${\cal E}_{\rm a}[\cdot]\approx U_{\rm t}[\cdot]U_{\rm t}^\dagger$. In particular in our simulations $F(\hat\chi,\chi_{\rm a})\gtrsim 0.95$ when  $F(\chi_{\rm t},\chi_{\rm a})\gtrsim 0.97$.  

However, as the implementation error increases and $F(\chi_{\rm t},\chi_{\rm a})$ decreases, the  performance of the three estimators depends  strongly on the nature of the errors.  The CS$_{\ell_1}$ is more sensitive to coherent errors than the CS$_\tr$ and LS estimators, as seen in Fig.~\ref{fig:fidFid}a.   Using the data from $d=5$ UIC input states, the fidelity between the CS$_{\ell_1}$ estimate and the applied map begins to fall below $\sim\!90\%$  in these simulations for $F(\chi_{\rm t},\chi_{\rm a})\lesssim 0.9$  while the CS$_\tr$ and LS estimators maintain their high fidelity. This trend is reversed for incoherent errors, as seen in Fig.~\ref{fig:fidFid}b. The CS$_{\ell_1}$ estimator  is more robust to incoherent errors of the form of Eq.~(\ref{depCh}) than either the CS$_\tr$ or LS estimators because the process matrix is no longer close to a low rank matrix, but it is still relatively close to a sparse matrix in the preferred basis.   As the incoherent error magnitude increases, the  CS$_{\ell_1}$ method returns an estimate with (on average) higher fidelity with the applied map than either the CS$_{\tr}$ or the LS estimates.  We thus conclude that when the applied map is sufficiently far from the target unitary map,  the performance of the three estimators varies in a manner that depends on the type of the error, and we can use this variation as an indicator of the type of error that occurred in the applied process. 

\begin{figure*}[t]
\centering
\includegraphics[width=1\textwidth]{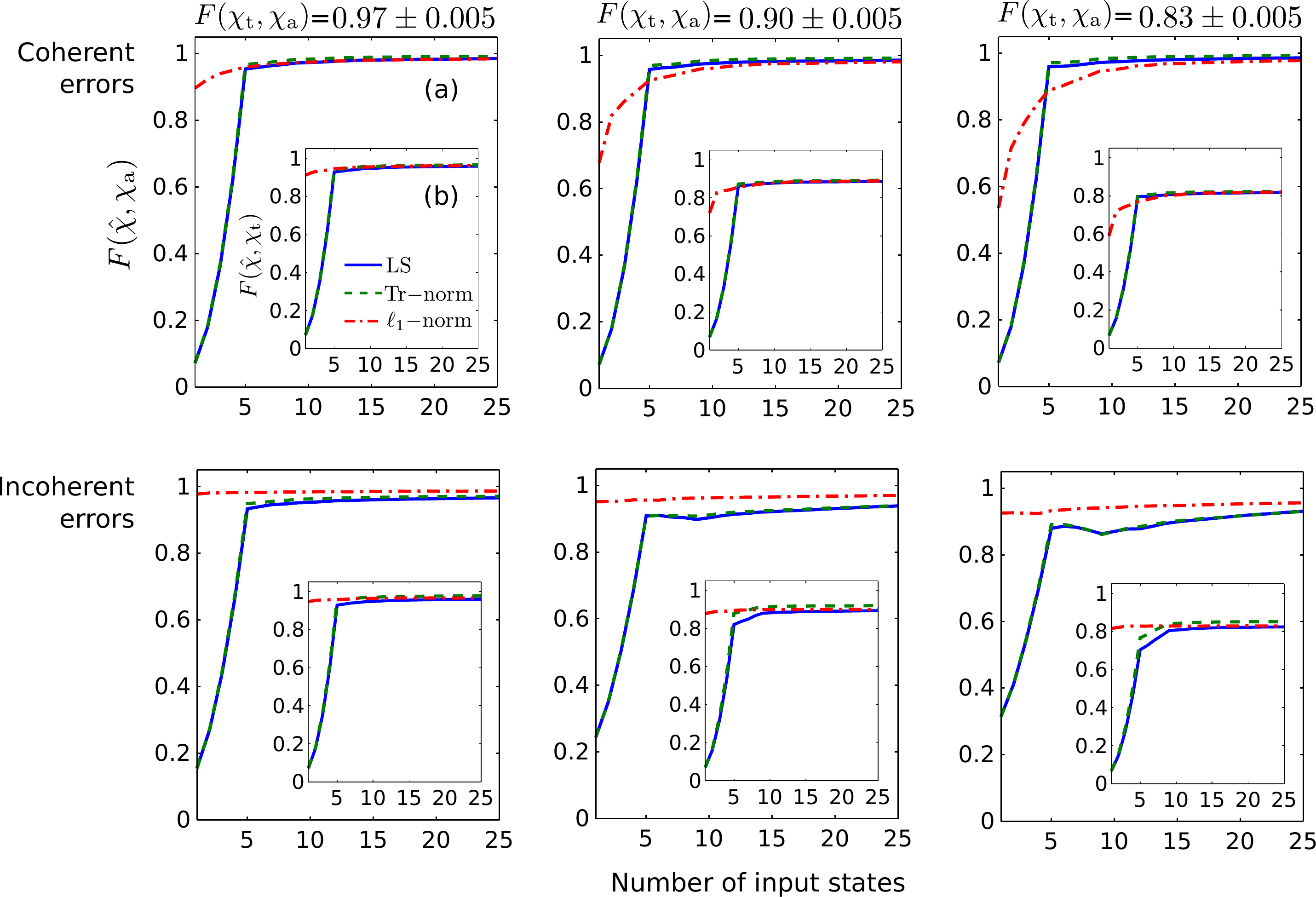} %
\caption{The fidelity between the estimate and the applied map, $F(\hat{\chi}, \chi_{\rm a})$ (a), and the estimate and the target map, $F(\hat{\chi}, \chi_{\rm t})$, inset (b), as a function of the number of input states, averaged over 20 applied processes, using different estimators, and under different error models for the applied map. Each column corresponds to a different magnitude of implementation error, represented by the fidelity between the applied and target map, $F(\chi_{\rm t},\chi_{\rm a})$. Top row: Coherent errors  as in Fig.~\ref{fig:fidFid}a. For a low error magnitude, $F(\chi_{\rm t},\chi_{\rm a})= 0.97\pm 0.005$, the three estimators return a high-fidelity estimate with the applied and the target maps. The CS$_ {\ell_1}$ estimator returns a reliable estimate with the information obtained from a single input state, while the CS$_ {\tr}$ and the LS minimization procedures  reliably estimates the map after five UIC input states. The CS$_ {\tr}$ and the LS estimators are robust to coherent error and perform essentially the same as the error  level increase,  here, $F(\chi_{\rm t},\chi_{\rm a})= 0.90\pm 0.005$ and $F(\chi_{\rm t},\chi_{\rm a})= 0.83\pm 0.005$ .  A noticeable slope in the fidelity as a function of the number of input states for the CS$_ {\ell_1}$ estimator, and a sharp kink in this curve for the LS and CS$_ {\tr}$ estimators, are indicators of the coherent error type.  Bottom row: Incoherent errors  as in Fig.~\ref{fig:fidFid}b. The CS$_ {\ell_1}$ estimator returns a reliable estimate with information  for higher error magnitudes, here, $F(\chi_{\rm t},\chi_{\rm a})= 0.90\pm 0.005$ and $F(\chi_{\rm t},\chi_{\rm a})= 0.83\pm 0.005$ with the fidelity relatively constant after a single input state.  In contrast, the CS$_ {\tr}$ and the LS estimators are more sensitive to an increase in the magnitude of incoherent errors.  As the error  level increases, the sharp transition in the fidelity as a function of the number of input states, becomes smoother as the noise increases.  These features of the fidelity curves for the three estimators are indicators of the incoherent error type.} 
 \label{fig:fidStates}
\end{figure*}

To understand the validation and diagnosis protocol, we study the fidelity between the estimate and the applied map (Fig.~\ref{fig:fidStates}a)  and between the estimate and the target map (Fig.~\ref{fig:fidStates}b) as a function of the number of input states, as studied previously in Sec.~\ref{sec:unitary}. The plots on the top and bottom rows correspond to different levels of coherent and incoherent errors. When the prior assumptions are valid and we are in a regime of a low error magnitude (either coherent or incoherent), e.g. $F(\chi_{\rm t},\chi_{\rm a})= 0.97\pm 0.005$ in these simulations, the three estimators yield reconstructions with high fidelity to the applied map (and with the target map). While the CS$_ {\tr}$ and the LS estimators require $d$ UIC input states to reliably characterize the applied map, with proper formulation, the  CS$_ {\ell_1}$ estimator returns a reliable estimate with information obtained from a single input state. 

As the error magnitude increases, our prior assumptions become less and less valid, and consequently the estimators yield lower fidelity with the data obtained from of order $d$ input states.  The data suggests that with high confidence the following conclusions hold.  First, the value of the fidelity $F(\hat\chi,\chi_{\rm t})$ obtained from $d$ UIC input states serves to validate that the applied map was close to a known target unitary map; the value of $F(\hat\chi,\chi_{\rm t})$  decreases when the applied map is further from the target.  Second, if the error in the applied map is not small, we can infer the dominant source of the imperfection by examining the behavior of the different estimators.  As seen in Fig.~\ref{fig:fidStates}a, with $F(\chi_{\rm t},\chi_{\rm a})= 0.83\pm 0.005$,  when employing  the CS$_ {\ell_1}$ estimator,  a large coherent error results in a curvature in $F(\hat\chi,\chi_{\rm t})$ as a function of the number of  input states.  Additionally, for the same data, using the CS$_ {\tr}$ and LS estimator, we see that  $F(\hat\chi,\chi_{\rm t})$ exhibits a sharp cusp after $d$ UIC probe states. In contrast, when the errors are dominantly incoherent, we see that when employing the CS$_ {\ell_1}$ estimator, $F(\hat\chi,\chi_{\rm t})$ is more or less a constant function of the number of input states.  In addition, there is a more gradual  increase of $F(\hat\chi,\chi_{\rm t})$ for the CS$_{\tr}$ and  LS estimators around $d$ states; the cusp behavior is smoothed.   These variations are signatures of the nature of the error in implementing the target unitary map.
 
In the regime $0.90\lesssim F(\chi_{\rm t},\chi_{\rm a})\lesssim 0.97$  it is difficult to distinguish, with high confidence, the nature of errors based solely on the behavior of $F(\hat\chi,\chi_{\rm t})$  as a function of input state,  and additional methods will be required to diagnose process matrix.  Nonetheless, a low fidelity of $F(\hat\chi,\chi_{\rm t})\lesssim 0.95$ after $d$ input states challenges the validity of our assumptions and indicates the presence of noise.

\section{Summary and conclusions}\label{sec:conc}
We have studied the problem of QPT under the assumption that the applied process is a unitary or close to a unitary map. We found that probing a unitary map on a $d$-level system with $d$ specially chosen pure input states (which we called UIC set of states) allows us to discriminate it from any other arbitrary CPTP map  given the corresponding output states.  In the ideal case of no errors, since the latter are completely characterized by a measurement of a POVM with a minimum of $2d$ elements, all together QPT of a unitary map requires measurement of a total of $2d^2$ POVM elements.  This is in comparison with $d^4-d^2$  POVM elements required for  discriminating  any two arbitrary CPTP map. We then showed that discriminating a unitary map from any other unitary map requires measurements of a minimum number of $d^2+d$ POVMs elements. 

We used the methods for efficient unitary map reconstruction to analyze a more realistic scenario where the applied map is close to a target unitary map and the collected data includes statistical errors. Under this assumption, we studied the performance of three convex--optimization--based estimators, the LS, CS$_\tr$ and CS$_ {\ell_1}$.  For each of these estimators we reconstructed the applied process from the same simulated data obtained by probing the map with pure input states, the first $d$ of which form a UIC set. We considered two types of errors that may occur on the target map, coherent errors, for which the applied map is a unitary map but slightly ``rotated'' from the target map, and incoherent errors in which the applied map is full rank but with high purity. In our simulation in Sec.~\ref{sec:noisy} we used the states of Eq.~(\ref{op basis nc}) to probe a randomly generated (applied) map with the desire properties. 

Our analysis suggests that when the prior assumptions are valid the three estimators yield high-fidelity estimates with the applied map using only the input  UIC set of states.  We found that the sensitivity of these methods for various types of errors yields important information about the validity of the prior assumptions and about the nature of the errors that occurred in the applied map. In particular, probing the map with UIC set of $d$ pure states and obtaining low fidelity between the estimates and the target map indicates that the errors are actually not small and the applied map is not close to the target unitary map. Furthermore, the performance of the different estimators and under coherent and incoherent noise, enables the identification of the dominant error type. One can then take this this information into account to further improve the implementation of the desire map.

Further extensions will be necessary before this protocol will be useful in practice.  While we have separately studied the effects of coherent and incoherent errors, in any real application both types of errors will occur to some degree.  We expect that if one error source sufficiently dominates, the signatures we found in the reconstruction that characterize the nature of the error will survive.  An additional extension of particular importance is a study of the performance of QPT in the presence of SPAM errors.  In the analysis presented here we assumed neither error in the preparation of the probe states nor in the POVM elements measured.  Errors of this sort will certainly contaminate the data and make it more difficult to both validate prior assumptions and diagnose errors.  Additional analysis is required to calibrate how SPAM errors affect the distinctive signatures of errors in gate implementations.

Finally, it will be important to study how the form the input states that form a UIC set affects our protocol.  For example the information learned from each input state of Eq.~(\ref{n+}) and Eq.~(\ref{0+n}) is qualitatively different, as seen in Fig.~\ref{fig:unitary_diff_sets}a. It is important to design the input states so that the information encoded in the correspondent output state would be as robust as possible.  The amplitude $1/{\sqrt2}$ of the states of Eq.~(\ref{0+n}) were chosen to ensure  robustness against statistical errors. For example, the set of states $\{\ket{0}, \sqrt{0.999}\ket{0}+\sqrt{0.001}\ket{n}, n=1,\ldots,d-1\}$  is UIC, but with large enough statistical errors the information encoded in the $\ket{n}$ states becomes noisy and uninformative. In future work we will examine how the choice of UIC states minimizes the error in the reconstruction due to statistical errors and perhaps, SPAM errors.

{\em Acknowledgments}. We gratefully acknowledge useful discussions with Poul Jessen, Carlos Riofr\'{\i}o, Matthew Grace, and Robert Kosut. This work was supported by NSF Grants PHY-1307520 and PHY-1212445.


\end{document}